\documentclass[preprint, 1p]{elsarticle}
\usepackage{graphicx} 
\usepackage{dcolumn}
\usepackage{tabularx}
\usepackage{booktabs}
\usepackage{hyperref}
\usepackage{boldline,multirow}
\usepackage{xcolor}
\usepackage{xspace}
\usepackage{listings}
\usepackage{accsupp}
\usepackage{caption}
\usepackage{subcaption}

\newcommand{\emptyaccsupp}[1]{\BeginAccSupp{ActualText={}}#1\EndAccSupp{}}

\newcommand\YAMLcolonstyle{\color{red}\mdseries}
\newcommand\YAMLkeystyle{\color{black}\bfseries}
\newcommand\YAMLvaluestyle{\color{blue}\mdseries}

\makeatletter

\newcommand\language@yaml{yaml}

\expandafter\expandafter\expandafter\lstdefinelanguage
\expandafter{\language@yaml}
{
  keywords={true,false,null,y,n},
  keywordstyle=\color{darkgray}\bfseries,
  basicstyle=\YAMLkeystyle,                                 
  sensitive=false,
  comment=[l]{\#},
  morecomment=[s]{/*}{*/},
  commentstyle=\color{purple}\ttfamily,
  stringstyle=\YAMLvaluestyle\ttfamily,
  moredelim=[l][\color{orange}]{\&},
  moredelim=[l][\color{magenta}]{*},
  moredelim=**[il][\YAMLcolonstyle{:}\YAMLvaluestyle]{:},   
  morestring=[b]',
  morestring=[b]",
  literate =    {---}{{\ProcessThreeDashes}}3
                {>}{{\textcolor{red}\textgreater}}1     
                {|}{{\textcolor{red}\textbar}}1 
                {\ -\ }{{\mdseries\ -\ }}3,
}

\lst@AddToHook{EveryLine}{\ifx\lst@language\language@yaml\YAMLkeystyle\fi}
\makeatother

\newcommand\ProcessThreeDashes{\llap{\color{cyan}\mdseries-{-}-}}

\newcommand{\cwb}{\texttt{cWB}\xspace}
\newcommand{\pycwb}{\texttt{PycWB}\xspace}
\newcommand{\cpp}{\texttt{C++}\xspace}

\title{PycWB: A User-friendly, Modular, and Python-based Framework for Gravitational Wave Unmodelled Search}
\author[1]{Yumeng Xu}
\author[1]{Shubhanshu Tiwari}
\author[2,3]{Marco Drago}

\affiliation[1]{organization={Physik-Institut, University of Zurich}, addressline={Winterthurerstrasse 190},
postcode={8057}, city={Zurich}, country={Switzerland}}

\affiliation[2]{Dipartimento di Fisica, Università di Roma ``La 
 Sapienza'', Piazzale Aldo Moro 2, I-00185 Roma, Italy
 }%
\affiliation[3]{INFN, Sezione di Roma, Piazzale Aldo Moro 2, I-00185 Roma, Italy
 }%

\begin{document}

\begin{abstract}    
    Unmodelled searches and reconstruction is a critical aspect of gravitational wave data analysis, requiring sophisticated software tools for robust data analysis. This paper introduces \texttt{PycWB}, a user-friendly and modular Python-based framework developed to enhance such analyses based on the widely used unmodelled search and reconstruction algorithm Coherent Wave Burst (\texttt{cWB}). The main features include a transition from C++ scripts to YAML format for user-defined parameters, improved modularity, and a shift from complex class-encapsulated algorithms to compartmentalized modules. The \texttt{PycWB} architecture facilitates efficient dependency management, better error-checking, and the use of parallel computation for performance enhancement. Moreover, the use of Python harnesses its rich library of packages, facilitating post-production analysis and visualization. The \texttt{PycWB} framework is designed to improve the user experience and accelerate the development of unmodelled gravitational wave analysis.
\end{abstract}

\maketitle

\section*{Metadata}

\begin{center}
\begin{table}[h!]
\begin{tabular}{ l l }
\toprule
    Current code version & 0.17.1 \\ 
\midrule
    Permanent link to code / repository & \href{https://git.ligo.org/yumeng.xu/pycwb}{\texttt{git.ligo.org/yumeng.xu/pycwb}} \\
\midrule
    Legal Code License & GNU General Public License \\
\midrule
    Code versioning system used & \texttt{git} \\
\midrule
    Software languages, tools, & \multirow{ 2}{*}{\texttt{Python}, \cpp, \texttt{Javascript}, \texttt{HTML}} \\
    and services used & \\
\midrule
    Compilation requirements & \multirow{ 2}{*}{\texttt{gcc}, \texttt{ROOT}, \texttt{wheel}, \texttt{setuptools}} \\
    and dependencies & \\
\midrule
    Link to developer & \multirow{ 2}{*}{\href{https://yumeng.xu.docs.ligo.org/pycwb}{\texttt{yumeng.xu.docs.ligo.org/pycwb}}} \\
    documentation / manual & \\
\midrule
    Support email for questions & \href{mailto:yumeng.xu@ligo.org}{\texttt{yumeng.xu@ligo.org}} \\
\bottomrule
\end{tabular}
\caption{Code metadata}
\label{table:metadata}
\end{table}
\end{center}

\section{Motivation and Significance}

The choice of programming language significantly influences the design and usage of scientific software. The benefits of having a Python-based software or Python interface for critical software in gravitational waves (GW) data analysis are outlined in \cite{SWIGLAL,pyseobnr}.  
Python, as of now, is on its way to becoming the default programming language in GW data analysis. This statement can be corroborated by the emergence of Python-based gravitational waveform models like \texttt{pySEOBNR}\cite{pyseobnr}, \texttt{gwsurrogate}\cite{gwsurrogate}, inference software like \texttt{BILBY}\cite{bilby}, \texttt{PyCBC-inference}\cite{pycbc_inference}. And the success and wide usage of GW data analysis algorithms like \texttt{PyCBC} \cite{pycbc_search}. 

Despite these advancements, there remain several opportunities where Python-based software can accelerate the usage and development of GW data analysis algorithms. One specific example is the so-called unmodelled search and waveform reconstruction algorithms in GW data analysis. The lack of readily available Python-based open-source software restricts the development and usage of un-modelled algorithms, limiting it primarily to researchers proficient in languages like C/C++ which is a low-level programming language. Creating Python-based solutions and interfaces will enhance participation and development in the field.

The Coherent Wave Burst (\cwb) algorithm has been at the forefront of advancements in GW astrophysics \cite{Klimenko:2015ypf}. The range of applicability of \cwb for GW transient data analysis is very wide as it is an all-sky morphology-independent algorithm i.e. it does not rely on the waveform models or the sky direction of the source. Instead \cwb relies on the coherent energy produced by the GW signal in the network of detectors. \cwb has played a major role in the discovery of the first detection of GW signal GW150914 \cite{GW150914} and more recently it has proved itself to be a crucial method to detect interesting transient GW signals that are not well modelled like GW190521 \cite{GW190521,Szczepanczyk:2020osv}. \cwb is routinely used in a variety of GW transient searches for the LIGO-Virgo-KAGRA collaborations like IMBH searches \cite{LIGOScientific:2021tfm}, eBBH searches \cite{LIGOScientific:2019dag} and generic searches for transients with short \cite{KAGRA:2021tnv} and long duration \cite{KAGRA:2021bhs}. 

While \cwb offers an extensive array of functionalities and scripts, however, it falls short in facilitating user-specific modifications not inherently supported by the framework. Although \cwb does provide plugin support, these plugins are required to access and manipulate global variables at the specific point of invocation. This approach demands a comprehensive understanding of the underlying code and risks unintentional disruptions or alterations to the variables. Moreover, the lack of clear dependencies between the modules further complicates the task for developers aiming to make modifications as the understanding of the interaction between different components becomes challenging. The \pycwb framework addresses these issues and offers a more straightforward and stable environment for customization and code alteration.

This paper introduces \pycwb, a modularized Python package for the \cwb algorithm. This package will enable the easier integration of the future machine learning algorithm and new Python-based waveform models. 
The remainder of this paper is structured as follows: Section \ref{sec:description} provides an introduction to the structure and features of \pycwb. Then, in Section \ref{sec:examples}, we present several use cases that demonstrate the user-friendliness and efficiency of \pycwb, comparing its application with the traditional \cwb. Finally, we share our conclusions and insights on the impact and potential of our new framework in the concluding section.

\section{Software Description}\label{sec:description}

The software framework in focus is implemented in Python and leverages the coherent Wave Burst (\cwb) software originally developed on \texttt{ROOT} \cite{ROOT}. The description of the core \cwb algorithm and the code can be found here \cite{cwb,klimenko_sergey_2021_5798976}. The native \texttt{pyROOT} interface of \texttt{ROOT} has immensely facilitated this Python implementation, saving the need for rewriting the entire suite of algorithms used for \cwb. Instead, the core \cwb code is integrated, specifically the \texttt{WAT} module, which is included in the package and automatically compiled upon installation using \texttt{pip}. The installation process is streamlined to avoid the usually intricate \cwb setup.

To install the \pycwb, we provide the easiest way which is to use conda due to its  dependencies on \texttt{ROOT} and \texttt{HEALPix}\cite{healpix} for \cwb core code

\begin{lstlisting}[
	language=Bash,
	label=lst_std,
	xleftmargin=\parindent,
	xrightmargin=0.7cm,
]
conda create -n pycwb "python>=3.9,<3.11"
conda activate pycwb
conda install -c conda-forge root=6.26.10 healpix_cxx=3.81 nds2-client python-nds2-client lalsuite setuptools_scm cmake pkg-config
python3 -m pip install pycwb
\end{lstlisting}

In its design, the software takes a modular approach. This way, the core \cwb code is divided into different modules, providing a roadmap for future transitions, where the existing C++ codes can be replaced seamlessly with Python modules.

\begin{figure}[h]\label{fig:pycwb_struct}
    \centering
    \includegraphics[width=\textwidth]{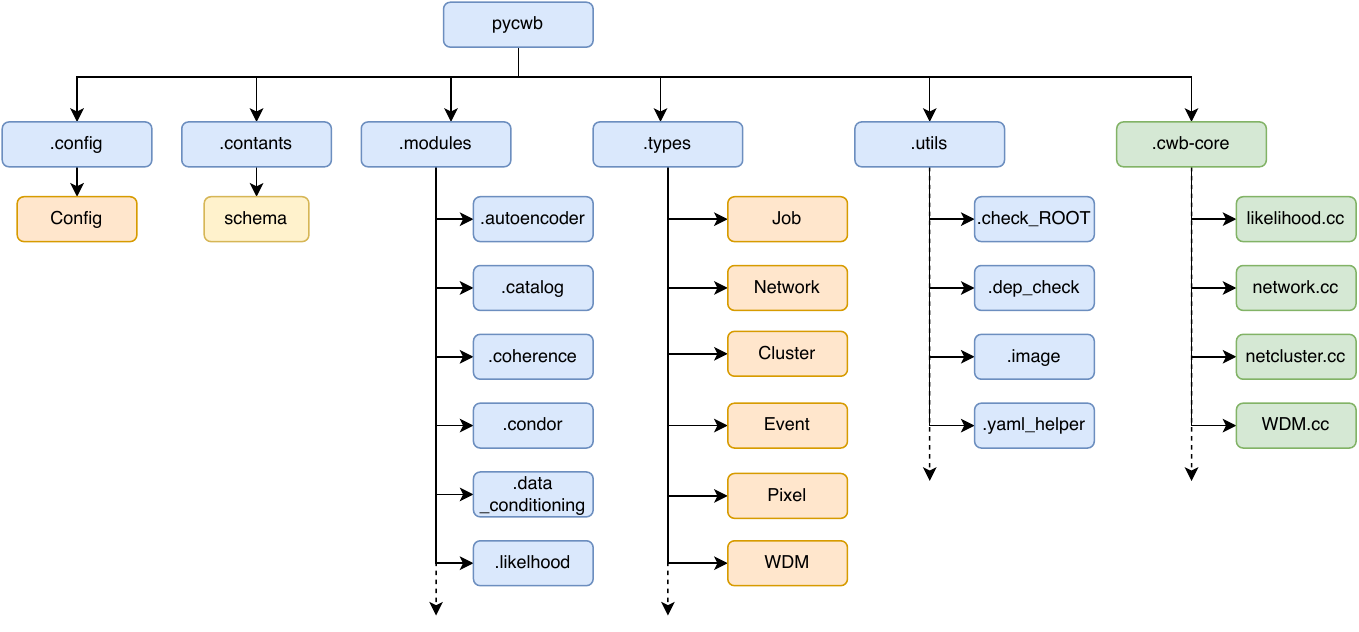}
    \caption{This figure shows the structure of the \pycwb. The blue blocks represent the Python modules and the orange blocks represent the Python class. The green blocks show the external C/C++ code embedded in \pycwb, while the yellow blocks highlight the key variables.}
    \label{fig:enter-label}
\end{figure}

\subsection{User Configurations}

The \pycwb framework makes it simpler for users to set their own parameters. Instead of using C++ scripts like before, \pycwb uses YAML format, a more user-friendly way that doesn't need C++ knowledge. Users or developers can easily set their own parameters using a Python dictionary. This dictionary is set up like a JSON schema, allowing flexibility in defining parameters. This system makes sure the user inputs are correct, checks them against the types and ranges that are already set, and provides default values when needed. Additionally it can automatically generate guides or instructions from this schema, keeping users updated with the software's requirements. This new approach makes it simpler for users to work with the software, manage their settings, and helps in faster development.

\noindent\begin{minipage}[t]{.45\textwidth}
\begin{lstlisting}[
caption=cWB user parameters,frame=tlrb,
language=C,
label=lst_std,
xleftmargin=\parindent,
xrightmargin=0.7cm,
]{Name}
  strcpy(analysis,"2G");

  nIFO = 3;
  cfg_search = 'r';	
  optim=false;

  strcpy(ifo[0],"L1");
  strcpy(ifo[1],"H1");
  strcpy(ifo[2],"V1");
  strcpy(refIFO,"L1");
  
  strcpy(channelNamesRaw[0],"L1:GWOSC-4KHZ_R1_STRAIN");
  strcpy(channelNamesRaw[1],"H1:GWOSC-4KHZ_R1_STRAIN");
  strcpy(channelNamesRaw[2],"V1:GWOSC-4KHZ_R1_STRAIN");

  strcpy(frFiles[0],"input/L1_frames.in");
  strcpy(frFiles[1],"input/H1_frames.in");
  strcpy(frFiles[2],"input/V1_frames.in");
\end{lstlisting}
\end{minipage}\hfill
\begin{minipage}[t]{.45\textwidth}
\begin{lstlisting}[
caption=PycWB user parameters,frame=tlrb,
language=Python,
label=lst_std,
xleftmargin=\parindent,
xrightmargin=0.7cm,
]{Name}
analysis: "2G"
cfg_search: "r"

optim: False

ifo: ["L1","H1","V1"]
refIFO: "L1"

channelNamesRaw: ['L1:GWOSC-4KHZ_R1_STRAIN', 'H1:GWOSC-4KHZ_R1_STRAIN', 'V1:GWOSC-4KHZ_R1_STRAIN']
frFiles: ["input/L1_frames.in", "input/H1_frames.in",  "input/V1_frames.in"]
\end{lstlisting}
\end{minipage}

\subsection{Modular and Classes}

The original \cwb framework presented a challenging structure where class-based constructions excessively encapsulated numerous methods and key algorithms. This approach led to dense coding, which was difficult to comprehend and modify due to its high interdependencies and complexity. However, in the \pycwb framework, a fundamental shift towards modularity and clarity is adopted. Essential functions are transferred to Python classes, serving as standard data formats or interfaces between different modules. Meanwhile, key algorithms have been detached from their original class environments and restructured into independent modules. This revamped architecture facilitates efficient dependency management. The necessary variables are initialized before each function call, and functions are called as pure procedures, thereby significantly enhancing the software's comprehensibility, usability, and adaptability.

\begin{figure}[h]\label{fig:pycwb_parallel}
    \centering
    \begin{subfigure}[b]{0.58\textwidth}
        \centering
        \includegraphics[width=\textwidth]{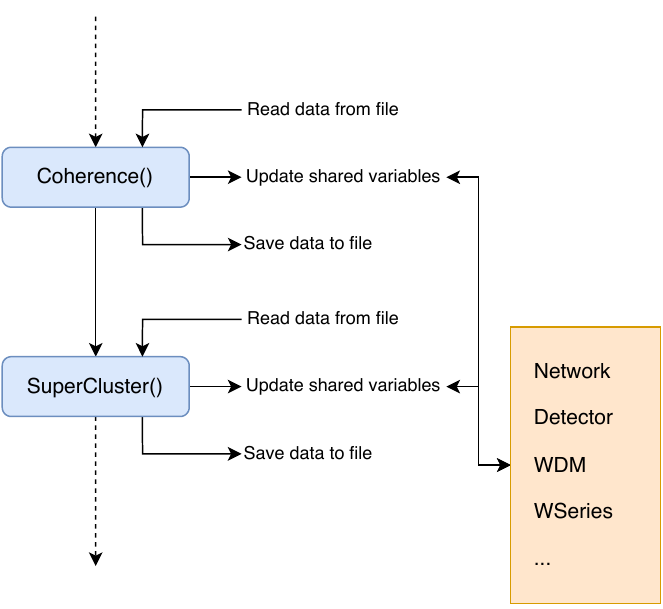}
        \caption{Part of the workflow for \cwb}
    \end{subfigure}
    \hfill
    \begin{subfigure}[b]{0.40\textwidth}
        \centering
        \includegraphics[width=\textwidth]{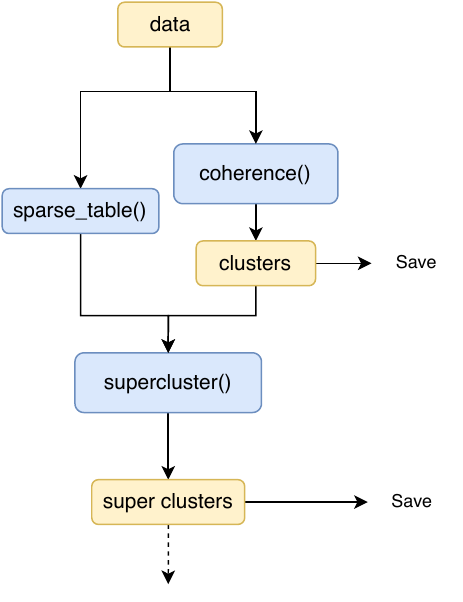}
        \caption{Part of the workflow for \pycwb}
    \end{subfigure}
    \caption{This figure shows the modular design of \pycwb. On the left, the workflow appears streamlined, but the updates on shared variables are scattered within the modules making it resemble a black box. On the right, we demonstrate our approach to modularization in \pycwb. We have isolated each module to ensure it only depends on the input and subsequently delivers the output to the next steps. This architecture enhances the transparency and flexibility of the process, allowing for easier comprehension and customization.}
    \label{fig:enter-label}
\end{figure}

\subsection{Parallelization}

The modularization in Python facilitates easy parallelization of various processes in \pycwb, using Python's \texttt{multiprocessing} library. As a result, computations across layers are expedited, bringing about a speedup of 4 to 6 times in the pixel finding and clustering stage. Reading data and data conditioning also enjoy speed improvements, leading to an overall speedup of 2 to 3 times. 

A further enhancement of performance by \pycwb can be envisaged by the integration of GPU acceleration. With the increase in the number of GW events and data the integration of GPU acceleration will become essential for \cwb algorithm. To this end \pycwb interface provides a much more straightforward solution with Numba\cite{Numba}.

\subsection{Post production processing}

With Python's wealth of packages for data processing, visualization, and machine learning, post-production data processing is seamless. The modular framework lets users easily select and implement the post-processing modules they need without requiring code modifications or recompilation. 
For instance, integrating the autoencoder neural network for glitch detection \cite{Bini:2023gil} into \pycwb is as simple as interfacing a few lines of TensorFlow\cite{TensorFlow} code with the framework. This ensures that advanced techniques such as machine learning can be employed efficiently and effectively without the need for extensive code modifications.

\begin{figure}[h]\label{fig:pycwb_postprod}
    \centering
    \includegraphics[width=0.8\textwidth]{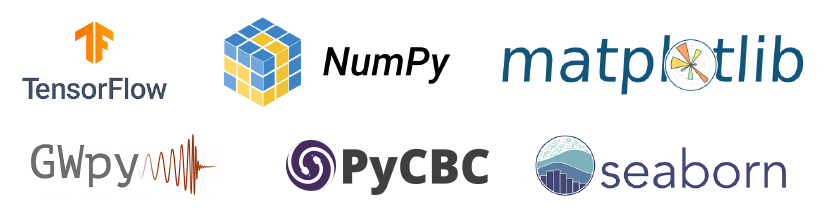}
    \caption{A selection of python packages that can be seamlessly integrated for post-production in \pycwb}
    \label{fig:enter-label}
\end{figure}

\subsection{Web interface}
Similar to \cwb, \pycwb also provides a web interface. But the web interface in \pycwb is structured as a separate module. This module contains HTML, CSS and Javascript frameworks as a web app, along with simple Python functions to copy the static webpage files to the designated output directory. As a result, there's no need for HTML webpage generation. This separation of the web interface and the Python code contributes to the modularity and usability of \pycwb.

\section{Illustrative Examples}\label{sec:examples}
\subsection{Real data analysis}

To validate the performance of \pycwb, we conducted identical analyses on real LIGO-Virgo events GW150914\cite{GW150914}, GW170809\cite{GWTC1}, and GW190521 \cite{GW190521} using both \cwb and \pycwb. For the \cwb analysis, we used the \texttt{cwb\_gwosc} command to download data from GWOSC \cite{GWOSC2,GWOSC3} and process them. 

\begin{lstlisting}[
	language=Bash,
	label=lst_std,
    numbers=none,
	xleftmargin=\parindent,
	xrightmargin=0.7cm,
]
 cwb_gwosc GW=EVENT_ID IFO=V1 TAG=TSTXY all
\end{lstlisting}

For \pycwb, we utilized the same user configuration file and data, executing the analysis via a simple command.

\begin{lstlisting}[
	language=Bash,
    numbers=none,
	label=lst_std,
	xleftmargin=\parindent,
	xrightmargin=0.7cm,
]
 pycwb_search user_parameters.yaml
\end{lstlisting}

Table \ref{table:perf_comp} outlines the events analyzed and the speed factors for both \cwb and \pycwb. The speed factor, calculated as the ratio of computation time to the length of the data, indicates a 2-3 times overall performance boost with \pycwb. This notable increase in speed is primarily attributed to effective parallelization made very simple to implement due to the Python interface.

\begin{center}
    \begin{table}[h]
    \begin{tabular}{ p{2.5cm} p{3cm} p{3cm} p{2cm}}
    \toprule
     \textbf{Event name} & \textbf{\cwb speed factor} & \textbf{\pycwb speed factor} & \textbf{relative improvement} \\ 
    \toprule
    GW150914 & 10X & 20.1X & 2 times \\
    \midrule
     GW170809 & 4.76X & 12.8X & 2.7 times \\
     \midrule
     GW190521 & 8.5X & 24.5X & 2.9 times \\
    \bottomrule
    \end{tabular}
    \caption{Performance comparison: \pycwb shows 2-3 times of performance improvement due to the parallelization compared to \cwb on the selected events.}
    \label{table:perf_comp}
    \end{table}
\end{center}

In addition, we highlight key parameters of the recovered events to showcase the consistency in accuracy between \cwb and \pycwb. Given that both platforms use the same algorithms and setups, similar accuracy levels are to be expected. Any minor deviations could potentially be attributed to differences in data types between Python and C++. These results affirm \pycwb's capabilities in maintaining the robustness of the algorithm while enhancing performance.

\begin{center}
    \begin{table}[h]
    \begin{tabular}{ p{2.5cm} l l l l}
    \toprule
     \textbf{Event name} & \textbf{Software} & \textbf{rho[0]} & \textbf{likelihood} & \textbf{start time}\\ 
    \toprule
    \multirow{ 2}{*}{GW150914} &\cwb & 16.70109 & 634.7065 & 1126259462.125 \\
    & \pycwb & 16.76488 & 641.5996 & 1126259462.125 \\
    \midrule
     \multirow{ 2}{*}{GW170809} &\cwb & 6.045537 & 117.5504 & 1186302519.6875  \\
      & \pycwb & 6.045362 & 117.5518 & 1186302519.6875 \\
     \midrule
     \multirow{ 2}{*}{GW190521} &\cwb & 10.13721 & 215.8701 & 1242442967.125 \\
      & \pycwb & 10.64826 & 236.0143 & 1242442967.125 \\
    \bottomrule
    \end{tabular}
    \caption{The results obtained from the \pycwb analysis are consistent with those obtained from \cwb, accommodating only the differences inherent in the data types between Python and C++. }
    \label{table:perf_comp}
    \end{table}
\end{center}

\begin{figure}
    \centering
    \begin{subfigure}[b]{0.57\textwidth}
        \includegraphics[width=\textwidth]{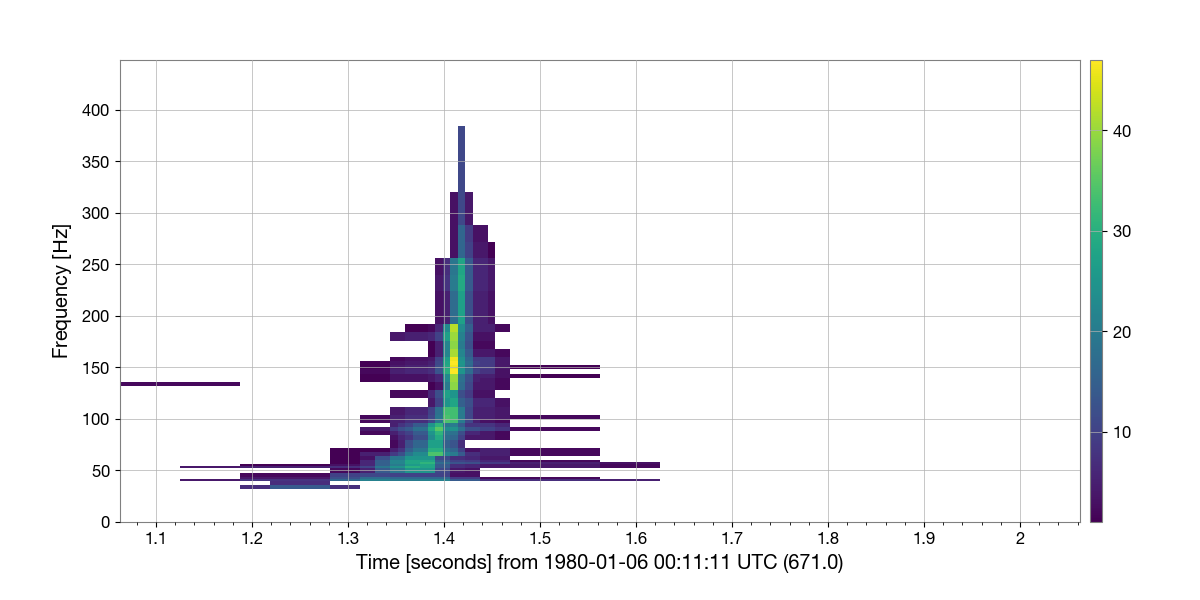}
        \caption{Likelihood map}
    \end{subfigure}
    \begin{subfigure}[b]{0.40\textwidth}
        \includegraphics[width=\textwidth]{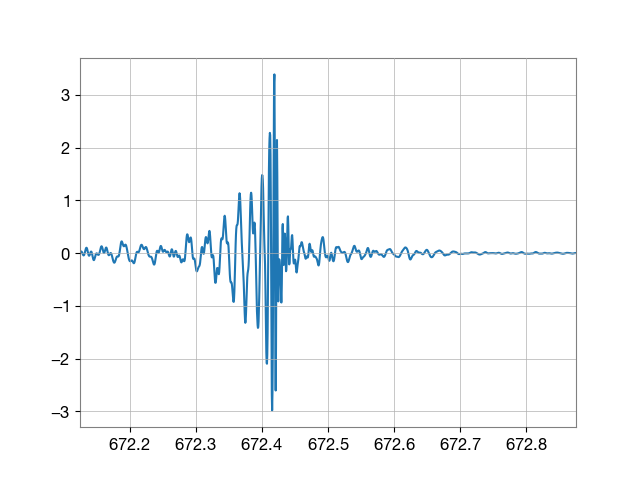}
        \caption{Reconstructed waveform}
    \end{subfigure}
    \caption{A selection of the plots from the \pycwb analysis on GW150914, the first detected GW event. These plots were created using Matplotlib, a popular data visualization library in Python, which takes the output from the \texttt{likelihood} module. In comparison with C++, Python's Matplotlib requires significantly less code for plotting, demonstrating the efficiency and ease of use provided by \pycwb in visualizing analysis results.}
    \label{fig:gw150914}
\end{figure}

Figure \ref{fig:gw150914} presents an example of an analysis conducted on GW150914 using \pycwb. The left panel displays the time-frequency map of the likelihood of selected pixels, providing a visual representation of how the event's signal changes over time and frequency. The right panel illustrates the reconstructed waveform of the event.

\subsection{Batch Injection with Python script}

The \pycwb framework simplifies the handling of batch injections involving large parameter sets. In \cwb, injecting complex GW signals (like binary black holes populations) is achieved with cumbersome scripts which create XML files, in particular the LIGO Light-Weight (LIGOLW) XML format \cite{ligo-lw-xml} for parameters. Moreover, one needs to modify the XML table manually when dealing with keys not predefined in the LIGOLW XML table. 
On the other hand, \pycwb provides the option to generate an array of parameters directly through a Python function. This function returns a list of parameters, offering significant flexibility and efficiency. Recently search algorithms like PyCBC etc provide HDF5 injection file support to mitigate the issues with LIGO-LW XML file formats, these injection files can be seemlessly integrated in \pycwb to have consistent injections between pipelines. 

The parameters are passed directly to the waveform generator, thus allowing users to employ their own waveform generators with additional parameters, circumventing the need for any code modification within \pycwb. The Python function can be constructed as demonstrated in the code snippet \ref{lst:generate_par}.

\begin{lstlisting}[
	language=Python,
	caption={generate\_parameters.py: A simple example for batch injection script},
	label={lst:generate_par},
	xleftmargin=\parindent,
	xrightmargin=0.7cm,
]
def get_injection_parameters():
    return [{
        'mass1': 20,
        'mass2': 20,
        'spin1z': spin1z,
        'spin2z': 0,
        'distance': 200,
        'inclination': 0,
        'polarization': 0,
        'gps_time': 1126259462.4,
        'coa_phase': 0,
        'ra': 0,
        'dec': 0
    } for spin1z in [0, 0.3, 0.6]]
\end{lstlisting}\label{lst:generate_parameters}

To implement this, the \texttt{parameters\_from\_python} keyword can be used in place of the \texttt{parameters} keyword in the YAML file containing user parameters. The example shows in code snippet \ref{lst:uf_generate_par}.

\begin{lstlisting}[
	language=yaml,
	caption={Example user parameter file for generating injection parameters from script},
        label={lst:uf_generate_par},
	xleftmargin=\parindent,
	xrightmargin=0.7cm,
]
injection:
  segment: # ....
  parameters_from_python:
    file: "generate_parameters.py"
    function: "get_injection_parameters"
  approximant: #...
  generator: #...
\end{lstlisting}

\subsection{Injection with third-party Python waveform models}
While \pycwb uses PyCBC as the default for waveform generation during injection, users may wish to utilize their own models or adopt new ones that have yet to be implemented in PyCBC. To accommodate this, \pycwb features an option within its \texttt{read\_data} module that allows for the integration of any waveform generator. The only requirement is for the user to write a wrapper function that accepts input from their injection parameters and returns the '+' (plus) and '×' (cross) polarization. This flexible approach enables seamless integration of a diverse array of waveform models, thereby expanding the analytical capabilities of the \pycwb framework.

We demonstrated this ability by running a data injection with the cutting-edge waveform model, \texttt{pyseobnr}. This model, only implemented in Python, utilizes the effective-one-body method and numerical relativity (NR) calibration, serving as a testament to the framework's compatibility and flexibility.

\begin{lstlisting}[
	language=Python,
	caption={Example code using the pySEOBNR waveform model for injection},
	label={lst:pyseobnr_wrapper},
	xleftmargin=\parindent,
	xrightmargin=0.7cm,
]
from pyseobnr.generate_waveform import GenerateWaveform

def waveform_generator(mass1, mass2, spin1x, spin1y, spin1z, spin2x, spin2y, spin2z, distance, inclination, polarization, coa_phase, f_lower, delta_t, **kwargs):
    parameters = {
        'mass1': mass1,
        'mass2': mass2,
        'spin1x': spin1x,
        'spin1y': spin1y,
        'spin1z': spin1z,
        'spin2x': spin2x,
        'spin2y': spin2y,
        'spin2z': spin2z,
        'distance': distance,
        'inclination': inclination,
        'polarization': polarization,
        'coa_phase': coa_phase,
        'f_ref': f_lower,
        'f22_start': f_lower,
        'deltaT': delta_t,
        "approximant": "SEOBNRv5HM",
    }
    wfm_gen = GenerateWaveform(parameters)  
    hp, hc = wfm_gen.generate_td_polarizations_conditioned_2()

    return hp, hc
\end{lstlisting}

To integrate a third-party waveform model that's not natively supported in \pycwb, the user only needs to write a wrapper function, as demonstrated in code snippet \ref{lst:pyseobnr_wrapper}. Then, to incorporate this into a run, the user simply adds a \texttt{generator} key to the user parameter file, specifying the location and function name, as shown in code snippet \ref{lst:pyseobnr_wrapper_yaml}.

\begin{lstlisting}[
	language=yaml,
	caption={Example for using the waveform model wrapper in user parameter file},
	label={lst:pyseobnr_wrapper_yaml},
	xleftmargin=\parindent,
	xrightmargin=0.7cm,
]
injection:
  segment: # ...
  parameters_from_python: # ...
  approximant: "SEOBNRv5HM"
  generator:
    module: "PATH_TO_THE_CODES"
    function: waveform_generator
\end{lstlisting}

\section{Conclusions}

The Coherent WaveBurst (cWB) is playing a key role in the discovery and analysis of gravitational waves. Despite its importance, the accessibility and user-friendliness of its interface have been a challenge, primarily due to the complex and highly technical nature of the C++ language it was originally written in.

In addressing this challenge, we present \pycwb, a Python-based modular adaptation of the cWB. This transformation not only makes cWB more accessible to the scientific community but also unlocks the potential for numerous innovations in the field. By leveraging the power and simplicity of Python, \pycwb makes complex analyses in gravitational wave research more manageable and user-friendly.
The capability to use \pycwb within interactive Jupyter Notebooks simplifies the learning process for new users, making it significantly more approachable. Moreover, the seamless compatibility with PyCBC and other Python-based waveform models such as \texttt{pyseobnr} and \texttt{gwsurrogate} greatly simplifies injection studies. 
Data post-processing is much easier. Its Python-based data output from each module makes data manipulation more intuitive. Additionally, users have the flexibility to choose the specific data they wish to save at each stage of the process. 

Furthermore, the \pycwb framework allows seamless integration with machine learning libraries in Python, paving the way for more sophisticated and automated analyses. This adaptability extends to developers, who can easily design new modules or add GPU-accelerated capabilities with Python packages like \texttt{numba}, without an in-depth understanding of the entire \pycwb structure.

In conclusion, \pycwb provides user-friendly, flexible, and powerful architecture, which opens new avenues for research and discovery in the realm of gravitational waves. The ease of use and adaptability of \pycwb empowers both experienced researchers and newcomers alike to contribute to this exciting field of research.

\section{Conflict of Interest}
We wish to confirm that there are no known conflicts of interest associated with this publication and there has been no significant financial support for this work that could have influenced its outcome.

\section*{Acknowledgements}

We are grateful to Giovanni Andrea Prodi, Francesco Salemi, Sophie Bini, Dixeena Lopez, Michael Ebersold, Edoardo Milotti and Gabriele Vedovato for their helpful comments and encouragement throughout this project.
Y. Xu is supported by China Scholarship Council.
S. Tiwari is supported by the Swiss National Science Foundation (SNSF) Ambizione Grant Number: PZ00P2-202204.
M. Drago acknowledges the support from the 
Amaldi Research Center funded by the MIUR program 
'Dipartimento di Eccellenza' (CUP:B81I18001170001) and 
the Sapienza School for Advanced Studies (SSAS).
The authors are grateful for computational resources provided by the LIGO Laboratory and supported by National Science Foundation Grants PHY-0757058 and PHY-0823459.
This paper has document number LIGO-P2300267.

This research has made use of data or software obtained from the Gravitational Wave Open Science Center (gwosc.org), a service of the LIGO Scientific Collaboration, the Virgo Collaboration, and KAGRA. This material is based upon work supported by NSF's LIGO Laboratory which is a major facility fully funded by the National Science Foundation, as well as the Science and Technology Facilities Council (STFC) of the United Kingdom, the Max-Planck-Society (MPS), and the State of Niedersachsen/Germany for support of the construction of Advanced LIGO and construction and operation of the GEO600 detector. Additional support for Advanced LIGO was provided by the Australian Research Council. Virgo is funded, through the European Gravitational Observatory (EGO), by the French Centre National de Recherche Scientifique (CNRS), the Italian Istituto Nazionale di Fisica Nucleare (INFN) and the Dutch Nikhef, with contributions by institutions from Belgium, Germany, Greece, Hungary, Ireland, Japan, Monaco, Poland, Portugal, Spain. KAGRA is supported by Ministry of Education, Culture, Sports, Science and Technology (MEXT), Japan Society for the Promotion of Science (JSPS) in Japan; National Research Foundation (NRF) and Ministry of Science and ICT (MSIT) in Korea; Academia Sinica (AS) and National Science and Technology Council (NSTC) in Taiwan.

\newpage

\appendix

\section{Full user parameter file example}
\begin{lstlisting}[
	language=yaml,
	caption={Run the search with python},
	label=lst_std,
	xleftmargin=\parindent,
	xrightmargin=0.7cm,
]
###### project ######

outputDir: "output"
nproc: 6
###### search ######

analysis: "2G"
cfg_search: "r"

optim: False

###### network configuration ######
ifo: ["L1","H1"]
refIFO: "L1"

# lags
lagSize: 1
lagStep: 1.
lagOff: 0
lagMax: 0

# superlags
slagSize: 1 # number of super lags (simulation=1) - if slagSize=0 -> Standard Segments
slagMin: 0
slagMax: 0
slagOff: 0

# job
segLen: 1200
segMLS: 600
segTHR: 200
segEdge: 10

# frequency
fLow: 16.
fHigh: 1024.


levelR: 3
l_low: 4 # low frequency resolution level		// std (sthr = 2)
l_high: 10 # high frequency resolution level	// std (sthr = 8)

wdmXTalk: "wdmXTalk/OverlapCatalog16-1024.bin"

healpix: 7

###### cWB production thresholds & regulators ######

bpp: 0.001
subnet: 0.5
subcut: -1.0
netRHO: 5.0
cedRHO: 5.0
netCC: 0.5
Acore: 1.7
Tgap: 0.2
Fgap: 128.0
delta: 0.5
cfg_gamma: -1.0
LOUD: 300

pattern: 10

iwindow: 100

# simulation
simulation: "all_inject_in_one_segment"
nfactor: 1

injection:
  segment:
    start: 1126258862.4
    end: 1126260062.4
    noise:
      seeds: [0, 1]
  parameters:
    mass1: 20
    mass2: 20
    spin1z: 0
    spin2z: 0
    distance: 500
    inclination: 0
    polarization: 0
    gps_time: 1126259462.4
    coa_phase: 0
    ra: 0
    dec: 0
  approximant: "IMRPhenomXHM"


\end{lstlisting}
\section{Example Python codes for \pycwb}

\begin{minipage}{\linewidth}
\begin{lstlisting}[
	language=Python,
	caption={Run the search with python},
	label=lst_std,
	xleftmargin=\parindent,
	xrightmargin=0.7cm,
]
from pycwb.search import search

search('./user_parameters.yaml')
\end{lstlisting}
\end{minipage}

\begin{lstlisting}[
	language=Python,
	caption={Run the search with python step by step},
	label=lst_std,
	xleftmargin=\parindent,
	xrightmargin=0.7cm,
]
import os

import pycwb
from pycwb.config import Config
from pycwb.modules.logger import logger_init

if not os.environ.get('HOME_WAT_FILTERS'):
    pyburst_path = os.path.dirname(os.path.abspath(pycwb.__file__))
    os.environ['HOME_WAT_FILTERS'] = f"{os.path.abspath(pyburst_path)}/vendor"

logger_init()

config = Config('./user_parameters_injection.yaml')

from pycwb.modules.read_data import generate_injection
from pycwb.modules.job_segment import create_job_segment_from_injection

# generate injected data for each detector with given parameters in config

job_segments = create_job_segment_from_injection(config.ifo, config.simulation, config.injection)

data = generate_injection(config, job_segments[0])

# apply data conditioning to the data
from pycwb.modules.data_conditioning import data_conditioning
from pycwb.modules.plot import plot_spectrogram

strains, nRMS = data_conditioning(config, data)

# calculate coherence
from pycwb.modules.coherence import coherence

fragment_clusters = coherence(config, strains, nRMS)

# supercluster
from pycwb.modules.super_cluster import supercluster
from pycwb.types.network import Network

network = Network(config, strains, nRMS)
pwc_list = supercluster(config, network, fragment_clusters, strains)

# likelihood
from pycwb.modules.likelihood import likelihood

events, clusters = likelihood(config, network, pwc_list) 
from pycwb.modules.plot import plot_event_on_spectrogram

# plot statistics
from gwpy.spectrogram import Spectrogram

for cluster in clusters:
    merged_map, start, dt, df = cluster.get_sparse_map("likelihood")
    plt = Spectrogram(merged_map, t0=start, dt=dt, f0=0, df=df).plot()
    plt.colorbar() 

from gwpy.spectrogram import Spectrogram

for cluster in clusters:
    merged_map, start, dt, df = cluster.get_sparse_map("null")
    plt = Spectrogram(merged_map, t0=start, dt=dt, f0=0, df=df).plot()
    plt.colorbar()
\end{lstlisting}

\bibliographystyle{apsrev4-1}
\bibliography{ref}

\end{document}